 \documentclass[final,5p,times,twocolumn,numbers]{elsarticle}

\usepackage{amssymb}
\usepackage{amsmath}
\usepackage{lipsum}

\journal{Physics Letters B}

\begin{document}

\begin{frontmatter}



\title{Black Holes, Warp Drives, and Energy Conditions}


\author[first]{Remo Garattini}
\author[first]{Kirill Zatrimaylov}
\affiliation[first]{organization={Università degli Studi di Bergamo, Dipartimento di Ingegneria e Scienze Applicate},
            addressline={Viale Marconi 5}, 
            city={Dalmine (Bergamo)},
            postcode={24044}, 
            state={},
            country={Italy}}

\begin{abstract}
Following the work of H. Ellis, we study warp drives in the gravitational field of a Schwarzschild black hole. We find that as long as the warp drive crosses the black hole horizon at a subluminal speed, the horizon would be effectively absent inside the warp bubble. Moreover, we discover that the black hole's gravitational field can alleviate the violations of the weak energy condition (WEC) and the null energy condition (NEC) and therefore decrease the amount of negative energy required to sustain a warp drive, which may be instrumental for creating microscopic warp drives in lab experiments. We also consider the thermodynamics of a warp bubble interacting with a black hole and point out some paradoxes that may indicate a gap in our understanding of them from the thermodynamic point of view.
\end{abstract}



\begin{keyword}
Black holes \sep Warp drives



\end{keyword}

\end{frontmatter}




\section{Introduction}\label{introduction}
The concept of warp drives was first introduced by Miguel Alcubierre in his seminal 1994 paper~\cite{Alcubierre:1994tu}, and elaborated upon by José Natario in~\cite{Natario:2001tk}, as well as in a multitude of other works~(\cite{Everett:1995nn}-\cite{Alcubierre:2017kqf}). A warp drive is a solution of General Relativity that has the appearance of a "bubble" propagating on some (flat or non--flat) spacetime background. The observers inside the bubble are in an inertial reference frame, which means warp drives do not require external energy sources to accelerate, and they may move at any speed (in principle including superluminal). This makes them a viable candidate for interstellar travel, but they have one significant downside: in order to sustain a bubble, one requires exotic matter with negative energy density.

As described by H. Ellis in~\cite{Ellis:2004aw}, the Schwarzschild metric, which describes a black hole, can be mapped to a warp drive--type metric with the use of a coordinate system known as Painlevé--Gullstrand coordinates, which makes it possible to embed a warp drive in a black hole background (a different, but somewhat similar proposal to directly combine a warp drive and a Schwarzschild black hole was put forward in~\cite{Schuster:2022ati}). In section~\ref{sec:Overview} of this paper, we briefly discuss the basics of Alcubierre warp drives. Then, in section~\ref{sec:BHWD}, we consider the situation when the warp drive is embedded in the exterior of a black hole (in the limit when the warp bubble is much smaller than the black hole, and moving in the radial direction towards it), and demonstrate that the black hole's gravitational field affects the warp drive energy conditions, making it possible to alleviate the violations of the weak energy condition (WEC) and the null energy condition (NEC). This implies that an external gravitational field can decrease the amount of negative energy required to sustain a warp drive, a fact which may be useful for the development of microscopic warp drive-like structures in a lab. We also show that as long as a warp drive that crosses the black hole horizon is subluminal, the horizon is effectively absent inside the warp bubble (i. e. both an incoming and an outgoing geodesic can pass through the patch of the horizon that is inside the bubble). We generalize these results to the case of arbitrary warp bubble size and direction in section~\ref{sec:Generic}. Then, in section~\ref{sec:Thermodynamics}, we consider the thermodynamics of a warp bubble crossing a black hole horizon. We conclude in section~\ref{sec:conclusions} with an overview of the paper's key results.

\section{A Brief Overview of Warp Drive Physics}\label{sec:Overview}
A warp drive metric, as defined by Natario in~\cite{Natario:2001tk}, is given by
\begin{equation}
\label{Natario}
-dt^2+\sum^3_{i=1}(dx^i+N^i(\vec{r},t)dt)^2
\end{equation}
in the "mostly plus" spacetime signature. In ADM variables, one can define it by setting the lapse function $N$ to $1$ and the inner metric $h_{ij}$ to $\delta_{ij}$.

The warp drive itself is a deformation of the metric localized in a bubble--shaped region that is moving on some (flat or non--flat) spacetime background. Its velocity is given by
\begin{equation}
v^i_s(t)=\frac{dx^i_s}{dt} \ ,
\end{equation}
where $x^i_s(t)$ is the position of the bubble's center. Assuming that the bubble is moving along the x--axis on a flat spacetime background, the metric would be given by
\begin{equation}
-dt^2+(dx-v_sf(r_s)dt)^2+dy^2+dz^2 \ ,
\end{equation}
which is known as the Alcubierre metric. Here $f(\vec{r}_s(t))$ is a function describing the shape of the warp, with $r_s$ given by
\begin{equation}
r_s(t) \ = \ \sqrt{(x-x_s(t))^2+y^2+z^2} \ .
\end{equation}
At small values of $r_s, f(r_s)=1$; then, after $r_s$ approaches some value $R$ (the size of the warp bubble), $f(r_s)$ steeply decreases to 0.

As was shown in~\cite{Alcubierre:1994tu} and~\cite{Alcubierre:2017kqf}, the Alcubierre warp drive violates both the weak energy condition (WEC) and the null energy condition (NEC). Namely, the former says that the inequality
\begin{equation}\label{EC}
T^{\mu\nu}n_\mu n_\nu \ \ge \ 0 \ ,
\end{equation}
where $T^{\mu\nu}$ is the stress-energy tensor, should hold for any timelike vector $n_\mu$. If we choose
\begin{equation}
n_\mu \ = \ \left(-1,0,0,0\right) \ ,
\end{equation}
the expression~\eqref{EC} is just the energy density of the warp field:
\begin{equation}\label{ED}
\rho \ = \ \frac{1}{16\pi G}\left(K^2-K_{ij}K^{ij}\right) \ .
\end{equation}
As the extrinsic curvature tensor for flat intrinsic metric $h_{ij}$ is given by
\begin{equation}
K_{ij} \ = \ \frac{1}{2}\left(\partial_iN_j+\partial_jN_i\right) \ ,
\end{equation}
the energy density is manifestly negative:
\begin{equation}
\rho \ = \ -\frac{f'^2v^2}{32\pi G}\left(\frac{y^2+z^2}{r^2_s}\right) \ ,
\end{equation}
so the WEC is explicitly violated.

Likewise, the NEC states that the condition~\eqref{EC} should hold for any \textit{null} vector $n_\mu$. As demonstrated in~\cite{Alcubierre:2017kqf}, if we consider a vector oriented along the x--axis, given by
\begin{equation}
n^\mu \ = \ \left(1,\pm1,0,0\right)
\end{equation}
in the orthonormal frame, then
\begin{eqnarray}
T_{\mu\nu}n^\mu n^\nu \ = \ \rho+T_{xx}\pm2T_{nx} \ ,
\end{eqnarray}
where
\begin{eqnarray}
T_{xx} \ = \ 3\rho \ = \ -\frac{3f'^2v^2}{32\pi G}\left(\frac{y^2+z^2}{r^2_s}\right) \ ,\\
T_{nx} \ = \ \frac{v}{16\pi G}\left(\partial^2_yf+\partial^2_zf\right) \ .
\end{eqnarray}
Now, if we average this expression over the direction of the vector, the last term vanishes, and we find
\begin{eqnarray}\label{ANEC}
\left<T_{\mu\nu}n^\mu n^\nu\right> \ = \ 4\rho \ ,
\end{eqnarray}
which means that NEC is also violated~\footnote{It is interesting to note that the ratio between $\rho$ and $T_{xx}$ is exactly 3. This is the same number appearing when one computes the radial pressure and the energy density in the Casimir effect~\cite{Garattini:2019ivd}.}.
\section{Warp Drive in Black Hole Field}\label{sec:BHWD}
Now, let us consider the more generic situation when the bubble propagates on some stationary \textit{non-flat} background. This means that the components of the shift vector $N^i$ have the form
\begin{equation}
N^i \ = \ (1-f(r_s))N^i_b(\vec{r})-f(r_s)v^i_s \ ,
\end{equation}
where $N^i_b(\vec{r})$ is the background metric.

In the particular case when the background metric is spherically symmetric, $N^i_b$ are given by
\begin{equation}
N^i_b(r) \ = \ \beta(r)\frac{x^i}{r} \ .
\end{equation}

In this case, the background metric can also be written in the more compact form in spherical coordinates
\begin{equation}\label{C1}
-dt^2+(dr-\beta(r)dt)^2+r^2d\Omega^2 \ .
\end{equation}
As shown by Painlevé and Gullstrand, the Schwarzshild metric
\begin{equation}
-\left(1-\frac{R_G}{r}\right)dt^2+\frac{dr^2}{1-\frac{R_G}{r}}+r^2d\Omega^2 \ ,
\end{equation}
where $R_G=2GM$ is the Schwarzschild radius, can be brought to the form~\eqref{C1} with 
\begin{equation}
\beta \ = \ \sqrt{\frac{R_G}{r}}
\end{equation}
via a coordinate transformation
\begin{equation}\label{PG}
t \ = \ T \ - \ \int \ dr \ \frac{\sqrt{\frac{R_G}{r}}}{1-\frac{R_G}{r}} \ .
\end{equation}
As suggested by Ellis in~\cite{Ellis:2004aw}, this relation can be used to embed an actual warp drive within the exterior of a black hole by replacing
\begin{equation}\label{C2}
N^i \ = \ \sqrt{\frac{R_G}{r}}\frac{x^i}{r} \rightarrow (1-f(r_s))\sqrt{\frac{R_G}{r}}\frac{x^i}{r}-f(r_s)v^i_s \ .
\end{equation}

Let us assume that the warp drive is moving along the x--axis:
\begin{equation}\label{Nx}
N^x \ = \ (1-f)\sqrt{\frac{R_G}{r}}\frac{x}{r}-fv \ ,
\end{equation}
and consider the limit in which the characteristic size of the warp drive (the support of the function $f$) is much smaller than the Schwarzschild radius of the black hole $R_G$. In this limit, $N_{y,z}\approx0$, and the energy density~\eqref{ED} reduces to
\begin{equation}
\rho \ \approx \ -\frac{1}{32\pi G}\left((\partial_yN^x)^2+(\partial_zN^x)^2\right) \ .
\end{equation}
From this, once again neglecting terms $\propto\frac{y}{r}$ and $\frac{z}{r}$, we obtain
\begin{equation}
-\frac{1}{32\pi G}\left(v+\sqrt{\frac{R_G}{r}}\right)^2f'^2\left(\frac{y^2+z^2}{r^2_s}\right) \ .
\end{equation}
Hence, for negative $v$ (i. e. the warp drive moving towards the black hole) with
\begin{equation}
|v|>\frac{1}{2}\sqrt{\frac{R_G}{r}}
\end{equation}
\textit{the amount of negative energy required to sustain a warp drive would be decreased by the black hole's gravitational field} (otherwise, or if the warp drive is moving in the opposite direction, it would be increased). This means that the external gravitational field can alleviate the violation of WEC in warp drive metrics.

Now, let us consider the null energy condition (NEC). Following~\cite{Santiago:2021aup}, we can write the other components of the stress--energy tensor: 
\begin{eqnarray}
T_{ni} \ = \ \frac{1}{8\pi G}\left(\partial_jK_{ij}-\partial_iK\right) \ = \ \frac{1}{16\pi G}\left(\partial^2N_i-\partial_i(\partial_jN^j)\right) \\
\notag
T_{ij} \ = \ \frac{1}{8\pi G}\left(N^k\partial_kK_{ij}+KK_{ij}+\partial_{[i}N_{k]}K_{kj}+\partial_{[j}N_{k]}K_{ki}-\right.\\
\left.-\delta_{ij}\left(N^k\partial_kK+\frac{1}{2}K^2+\frac{1}{2}K_{kl}K^{kl}\right)+\partial_t\left(\delta_{ij}K-K_{ij}\right)\right) \ .
\end{eqnarray}
The latter can also be cast in the form
\begin{eqnarray}\label{SE}\notag
T_{ij} \ = \ \frac{1}{8\pi G}\partial_k\left(N^k\left(K_{ij}-K\delta_{ij}\right)\right)+\frac{\delta_{ij}}{16\pi G}\left(K^2-K_{kl}K^{kl}\right)+\\
+\frac{1}{16\pi G}\left(\partial_iN^k\partial_jN^k-\partial_kN_i\partial_kN_j\right)+\frac{1}{8\pi G}\partial_t\left(\delta_{ij}K-K_{ij}\right) \ ,
\end{eqnarray}
where the first term is just a total divergence, and the second term is the energy density $\rho$, multiplied by $\delta_{ij}$. Specifically, we get
\begin{eqnarray}
T_{nx} \ = \ -\frac{1}{16\pi G}\left(\partial^2_yN^x+\partial^2_zN^x\right) \ ,\\
T_{xx}  \ = \ -\frac{3}{32\pi G}\left((\partial_yN^x)^2+(\partial_zN^x)^2\right) \ = \ 3\rho \ .
\end{eqnarray}

This means that for a vector oriented along the x--axis, the averaged expression $\left<T_{\mu\nu}n^\mu n^\nu\right>$ would once again yield $4\rho$, as in~\eqref{ANEC}, so the violation of NEC would also be alleviated by the black hole's gravitational field.

As shown in~\cite{White:2021hip}, it may be possible to create warp drive--like structures within Casimir cavities in a lab, so it appears a promising direction to investigate how they would be affected by an external gravitational field (and whether such a field can serve a practical purpose by making it possible to realize a warp drive with a lesser amount of negative energy -- or, vice versa, to amplify the effect of a warp drive).

\begin{figure}
	\centering 
	\includegraphics[width=0.4\textwidth]{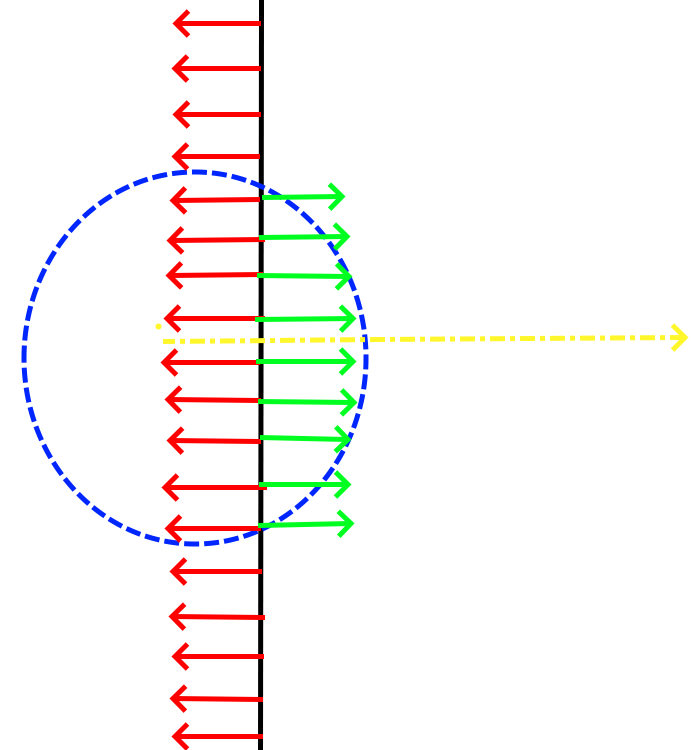}	
	\caption{Warp bubble (blue dashed line) crossing the black hole horizon (black line). Outside the bubble, there are only incoming null geodesics (red arrows), while inside the bubble, there are also outgoing ones (green arrows), making it possible for an observer inside the bubble to send a light signal to the outside from behind the horizon (yellow dash--dotted line).} 
	\label{fig_mom0}
\end{figure}

Finally, let us note that, as the $00$--component of the metric tensor is approximately
\begin{equation}
g_{00} \ \approx \ -1+\left((1-f)\sqrt{\frac{R_G}{r}}+f|v|\right)^2 \ ,
\end{equation}
\textit{the horizon would be effectively absent inside the warp bubble as long as it is subluminal} ($|v|<1$). One can understand this more precisely by considering a light ray moving along the x--axis, with the geodesic given by:
\begin{equation}
-dt^2+\left(dx+N^xdt\right)^2 \ \approx \ 0
\end{equation}
(here we once again neglected the term $((N^y)^2+(N^z)^2)dt^2$). Dividing this expression by $dt^2$, we obtain the quadratic equation
\begin{equation}
\left(\frac{dx}{dt}\right)^2+2N^x\left(\frac{dx}{dt}\right)+\left((N^x)^2-1\right) \ = \ 0
\end{equation}
with the roots
\begin{equation}
\frac{dx}{dt} \ = \ -N^x\pm1 \ .
\end{equation}
The plus and minus sign correspond to the outgoing and incoming geodesic respectively. At the Schwarzschild radius, i. e.
\begin{equation}
r=R_G \ ,
\end{equation}
the roots are given by 
\begin{equation}\label{Out}
\frac{dx}{dt} \ = \ f(v+1)
\end{equation}
and
\begin{equation}\label{In}
\frac{dx}{dt} \ = \ f(v+1)-2 \ .
\end{equation}
The outgoing geodesic (\eqref{Out}) would not vanish for $v>-1$, while the incoming one (\eqref{In}) does not vanish as long as $v<1$. Hence, for $|v|<1$, we would have both an outgoing and an incoming geodesic passing through the intersection area between the warp bubble and the black hole horizon (fig.~\ref{fig_mom0}). This means that an observer inside the warp bubble that has crossed the horizon can nonetheless send a light signal out of the black hole.

\section{Generic Warp Drives}\label{sec:Generic}
We may also consider the generic case when the warp drive is not necessarily much smaller than the black hole, and not necessarily moving in the radial direction (i. e. $N^{y,z}$ are non--negligible). Once again following~\cite{Santiago:2021aup}, one can write the energy density in the form
\begin{equation}\label{rho}
\rho \ = \ \frac{1}{16\pi G}\left(\partial_i(N_i\partial_jN_j-N_j\partial_jN_i)-\frac{1}{4}(\partial_iN_j-\partial_jN_i)^2\right) \ .
\end{equation}
Now, if we take
\begin{equation}
N^i \ = \ (1-f)\sqrt{\frac{R_G}{r}}\frac{x^i}{r}-fv^i \ ,
\end{equation}
the first term vanishes, and the second term becomes
\begin{equation}
-\frac{f'^2}{32\pi Gr^2_s}\left|\left(\vec{v}+\sqrt{\frac{R_G}{r}}\frac{\vec{r}}{r}\right)\otimes\vec{r}_s\right|^2 \ = \ -\frac{f'^2\sin^2\theta}{32\pi G}\left|\vec{v}+\sqrt{\frac{R_G}{r}}\frac{\vec{r}}{r}\right|^2 \ .
\end{equation}
The condition for the modulus of this expression to be decreased is
\begin{equation}
v^2+2v\sqrt{\frac{R_G}{r}}\cos\psi+\frac{R_G}{r} \ < \ v^2 \ ,
\end{equation}
where $\psi$ is the angle between the vectors $\vec{r}$ and $\vec{v}$. The violation of WEC is therefore alleviated under the condition
\begin{equation}\label{Cond}
v\cos\psi \ < -\frac{1}{2}\sqrt{\frac{R_G}{r}} \ ,
\end{equation}
i. e. the projection of $\vec{v}$ on the $r$--axis should be negative and greater by modulus than $\frac{1}{2}\sqrt{\frac{R_G}{r}}$. From this we automatically get the weaker condition
\begin{equation}
|v| \ > \frac{1}{2}\sqrt{\frac{R_G}{r}} \ .
\end{equation}

Now, let us consider NEC. Following~\eqref{SE}, we can write the spatial components of the stress--energy tensor as
\begin{eqnarray}
T_{ij} \ = \ \rho\delta_{ij}+\frac{f'^2}{16\pi G}\left(\frac{(x-x_s)_i(x-x_s)_j}{r_s^2}\vec{V}^2-V_iV_j\right)
\end{eqnarray}
up to the total divergence term and the time--derivative term that is first--order in $(x-x_s)$, and hence also vanishes upon integration. Here
\begin{equation}
V_i \ = \ v_i+\sqrt{\frac{R_G}{r}}\frac{x_i}{r} \ .
\end{equation}
Once again, we can assume without loss of generality that the velocity vector $\vec{v}$ is oriented along the x--axis. Therefore we would have
\begin{eqnarray}\notag
T_{xx} \ = \ \rho+\frac{1}{16\pi G}\left(\frac{R_G}{r}\left(\frac{y^2+z^2}{r^2}\right)(\partial_xf)^2-\right.\\
\left.-\left(v+\sqrt{\frac{R_G}{r}}\frac{x}{r}\right)^2\left((\partial_yf)^2+(\partial_zf)^2\right)\right) \ .
\end{eqnarray}
If the condition~\eqref{Cond} is satisfied, the first term with $\rho$ and the third term with $\partial_yf$ and $\partial_zf$ (both of them negative) would both be decreased by the gravitational field compared to a "free" warp drive. Besides, as long as $\frac{y}{r}$ and $\frac{z}{r}$ are non--negligible, we also have the second term, which is positive--definite, and hence also alleviates the violation of NEC.

This result is particularly useful given that the Schwarzschild metric describes not just black holes, but also generic spherically symmetric masses. While it's not possible to create a black hole in a lab, one could study the behavior of Casimir plates in the vicinity of a spherical or pointlike gravitational source.

Likewise, we can generalize the other result of this paper regarding the warp bubbles crossing black hole horizons. For non--negligible $N^{y,z}$, the equation for a light ray along the x--axis is given by
\begin{equation}
\left(\frac{dx}{dt}\right)^2+2N^x\left(\frac{dx}{dt}\right)+\left(\vec{N}^2-1\right) \ = \ 0 \ ,
\end{equation}
with the roots
\begin{equation}
\frac{dx}{dt} \ = \ -N^x\pm\sqrt{(N^x)^2+1-\vec{N}^2} \ .
\end{equation}
The outgoing geodesic, corresponding to the "+" sign, vanishes at $\vec{N}^2=1$, so we have to demand
\begin{equation}
\vec{N}^2(R_G) \ = \ (1-f)^2+f^2v^2-2(1-f)fv\cos\psi \ < \ 1 \ ,
\end{equation}
or, equivalently,
\begin{equation}
f(1+v^2)-2-2(1-f)v\cos\psi \ < \ 0 \ .
\end{equation}
For $f=1$, the equation is satisfied if $|v|<1$, which also remains valid for very small $f$. This is exactly the same condition as the one we obtained in the simplified case.

\section{Thermodynamics}\label{sec:Thermodynamics}
One would also be tempted to ask how does the warp bubble affect the black hole's thermodynamics. Normally, the Hawking temperature associated with horizons is proportional to surface gravity $\kappa$:
\begin{equation}
T_H \ = \ \frac{\kappa}{2\pi} \ ,
\end{equation}
which, in turn, is expressed through the timelike Killing vector $\xi$:
\begin{equation}\label{SG}
\xi^\beta\nabla_\beta\xi^\alpha \ = \ \kappa\xi^\alpha \ .
\end{equation}
However, the timelike Killing vector is only defined for stationary spacetimes, and in our case, it's manifestly dynamical. Though there have been attempts to generalize the notion of surface gravity to dynamical metrics, there is still no universal definition~\cite{Nielsen:2007ac,Pielahn:2011ra}. It is also not possible to switch to the warp bubble's rest frame, as was done in~\cite{Hiscock:1997ya,Finazzi:2009jb}, because in this frame, the black hole horizon would instead be moving. The only reasonable approximation we can make is that the warp bubble is moving very slowly ($v<<1$), and the metric can be considered stationary up to corrections of order $v^2$. This approximation also makes sense as an approximation of thermodynamic equilibrium~\footnote{In principle, we can also consider stationary warp bubbles, for which the function $f$ is time--independent: in this case, the prefactor $v$ in front of $f$ no longer has the physical interpretation of velocity and may instead be understood as just the measure of the distortion of spacetime.}.

Taking
\begin{equation}
\xi^\alpha \ = \ \left(1 \ , \ 0 \ , \ 0 \ , \ 0\right) \ ,
\end{equation}
we obtain from~\eqref{SG}
\begin{equation}
\kappa \ = \ \Gamma^0_{00} \ = \ -\frac{1}{2}g^{0i}\partial_ig_{00} \ .
\end{equation}

In the simplest case, when the warp bubble is much smaller than the black hole horizon, this yields
\begin{equation}
\kappa \ = \ -(N^x)^2\partial_xN^x \ ,
\end{equation}
which at horizon becomes
\begin{equation}
-(1+v)f'
\end{equation}
(here $f'$ is computed at the characteristic size of the warp bubble where $f\approx0$, and we neglected a term of order $R^{-1}_G$). The temperature is therefore
\begin{equation}
T \ = \ -\frac{1+v}{2\pi}f' \ .
\end{equation}

This result generalizes quite simply to the case when the warp bubble's velocity has a component tangential to the horizon: in this case, we have
\begin{equation}
N^x \ = \ \left(1-f\right)\sqrt{\frac{R_G}{r}}-fv_x \ , \ N^y \ = \ -fv_y \ .
\end{equation}
Since $N^x=1$ and $N^y=0$ at the horizon, the temperature would still be
\begin{equation}
T \ = \ -\frac{1}{2\pi}\partial_xN^x \ = \ -\frac{1+v_x}{2\pi}f'
\end{equation}

Now, let us consider the issue of entropy. Under the assumption that it's still proportional to the area of the horizon, the change in entropy induced by the warp bubble crossing the horizon would be given by
\begin{equation}\label{Entropy}
\delta S \ = \ \frac{1}{4l^2_P}\left(A_{cap}-A_{base}\right) \ .
\end{equation}
where
\begin{equation}
A_{cap} \ = \ 2\pi r_W\left(r_W-\delta\right)
\end{equation}
is the area of the spherical cap that is the part of the warp bubble behind the black hole horizon, and
\begin{equation}
A_{base} \ = \ \pi\left(r^2_W-\delta^2\right)
\end{equation}
is the area of the base of the cap that is also the part of the black hole horizon "cut out" by the warp bubble (here $r_W$ is the radius of the warp bubble, and $\delta<R_W$ is the distance between the center of the bubble and the black hole horizon). As
\begin{equation}
A_{cap} \ > \ A_{base} \ ,
\end{equation}
the positive first term in~\eqref{Entropy} would be greater than the negative second, and the overall change in entropy would be positive, which means that the warp bubble is amenable to black hole thermodynamics.

However, it remains unclear what would happen once the bubble is completely "swallowed" by the black hole, as it's an object with negative energy density, meaning that, at face value, it should \textit{decrease} the mass of the black hole and, therefore, the entropy. Matters are further complicated if we consider arbitrary warp bubbles that can be comparable to or larger in size than the black hole. Let us consider a simple Gedankenexperiment and assume that the warp bubble is \textit{much} larger than the black hole. In this case, instead of the warp bubble crossing the black hole horizon, we would have the black hole passing through the warp bubble. While the black hole is at the center of the bubble, its gravitational field would still be "felt" by distant observers, but the horizon would be effectively absent. This means that entropy (at least, in the original Hawking sense) can no longer be defined, and leads to other potential paradoxes.

Assuming that warp drives are not a physical impossibility and can be realized in nature, this result would imply a fundamental gap in our understanding of these objects. One possible solution is to attribute entropy and thermodynamic degrees of freedom to the warp bubble itself, even when it's sublumunal and hence possesses no horizon. However, such considerations go far beyond the scope of this paper.

\section{Conclusions}\label{sec:conclusions}
In this paper, we studied the model of warp drive embedding into the exterior of a black hole, following the scheme proposed by Ellis in~\cite{Ellis:2004aw}~\footnote{Despite superficial similarity, this line of research is drastically different from the one proposed in~\cite{Schuster:2022ati} -- in~\cite{Ellis:2004aw} we have an ordinary Alcubierre warp drive interacting with a black hole, while~\cite{Schuster:2022ati} introduces a modified warp drive that \textit{itself} has the properties of a Schwarzschild black hole.}. We discovered that the black hole's gravitational field would affect the stress--energy tensor of the warp drive, and that, under certain circumstances (namely, if the warp drive is moving towards the black hole, and its velocity is large enough), it can alleviate the violations of the null energy condition (NEC) and the weak energy condition (WEC) and reduce the amount of negative energy density required to sustain a warp drive. This finding bears some intriguing analogues with the Schwinger effect in quantum field theory that involves the conversion of virtual particles into real ones in an external electric field (see~\cite{Schwinger:1951nm}), and may also be instrumental for creating microscopic warp drives within Casimir cavities, such as the one described in~\cite{White:2021hip}.

Besides, we pointed out that as long as a warp bubble is subluminal, it can effectively "remove" the black hole horizon. Namely, both an incoming \textit{and an outgoing} geodesic can pass through the warp bubble's intersection area with the horizon. Interestingly, if a warp bubble is going out of a black hole at superluminal speed, there is a "white hole" effect: i. e. only the outgoing geodesic remains in the intersection region, and the incoming one vanishes.

Finally, we analyze black hole thermodynamics in the presence of a warp drive crossing the horizon and find that, assuming certain approximations (namely, that the warp bubble is much smaller than the black hole horizon, and is moving very slowly) and the validity of the entropy area law, the warp bubble would produce an increase in the black hole's entropy. However, there are potential problematic issues in other physical situations: namely, when the warp drive is completely absorbed by the black hole, it may decrease its mass, and, therefore, its entropy. Likewise, when there is a larger warp bubble passing through a black hole, it would produce a "screening" effect and de facto eliminate the horizon, making it impossible to define the black hole entropy in the Hawking sense. If warp drives are possible in nature, these issues indicate that we still do not understand them from the thermodynamic point of view.

\section{Declaration of Competing Interest}
The authors declare the following financial interests/personal relationships which may be considered as potential competing interests: Remo Garattini reports financial support was provided by Limitless Space Institute. Kirill Zatrimaylov reports financial support was provided by Limitless Space Institute. Remo Garattini reports a relationship with Limitless Space Institute that includes: funding grants. Kirill Zatrimaylov reports a relationship with Limitless Space Institute that includes: funding grants. If there are other authors, they declare that they have no known competing financial interests or personal relationships that could have appeared to influence the work reported in this paper.

\section{Acknowledgements}
We are grateful to Harold "Sonny" White for the discussions and for his useful feedback on earlier versions of this work, and to Prof. Claudio Maccone for his instructive comments and questions. The work is supported by the 2023 LSI grant "Traversable Wormholes: A Road to Interstellar Exploration" of Texas A \& M Engineering Experiment Station (TEES), The Texas A \& M University System. LSI-GF-2022-00. Part of the computations in this work was done with OGRe, a General Relativity Mathematica package developed by Barak Shoshany~\cite{Shoshany:2021iuc}.

\section{Data availability}
No data was used for the research described in the article.

\end{document}